\documentclass{llncs}
\usepackage{makeidx}  %
\usepackage[utf8]{inputenc} %
\usepackage[T1]{fontenc}    %
\usepackage{hyperref}       %
\usepackage{url}            %
\usepackage{booktabs}       %
\usepackage{amsmath, amsfonts}       %
\usepackage{nicefrac}       %
\usepackage{microtype}      %
\usepackage{bm}
\usepackage[%
algoruled, %
]{algorithm2e}
\usepackage[separate-uncertainty = true]{siunitx}
\usepackage{nicefrac}
\usepackage{pgfplots}
\usepackage{subcaption}
\usepackage{graphicx}
\usepackage{float}

\newlength{\figureheight}
\newlength{\figurewidth}
\DeclareMathOperator*{\argmax}{arg\,max}

\begin{document}
	
	\title{Guided Deep Reinforcement Learning for Swarm Systems}
	
	\author{Maximilian Hüttenrauch\inst{1} \and Adrian Šošić\inst{1} \and Gerhard Neumann \inst{2}}
	
	\institute{TU Darmstadt, Darmstadt, Germany
		\and
		University of Lincoln,
		Lincoln, UK}
	
	\maketitle
	
	\begin{abstract}
		In this paper, we investigate how to learn to control a group of cooperative agents with limited sensing capabilities such as robot swarms. The agents have only very basic sensor capabilities, yet in a group they can accomplish sophisticated tasks, such as distributed assembly or search and rescue tasks. Learning a policy for a group of agents is difficult due to distributed partial observability of the state. Here, we follow a guided approach where a critic has central access to the global state during learning, which simplifies the policy evaluation problem from a reinforcement learning point of view. For example, we can get the positions of all robots of the swarm using a camera image of a scene. This camera image is only available to the critic and not to the control policies of the robots. We follow an actor-critic approach, where the actors base their decisions only on locally sensed information. In contrast, the critic is learned based on the true global state. Our algorithm uses deep reinforcement learning to approximate both the Q-function and the policy. The performance of the algorithm is evaluated on two tasks with simple simulated 2D agents: 1) finding and maintaining a certain distance to each others and 2) locating a target.
	\end{abstract}
	
	
	\section{Introduction}
	\label{sec:intro}
	\noindent Nature provides many examples where the performance of a collective of limited beings exceeds the capabilities of one individual. Ants transport prey of the size no single ant could carry, termites build nests of up to nine meters in height, and bees are able to regulate the temperature of a hive. Common to all these phenomena is the fact that each individual has only basic and local sensing of its environment and limited communication capabilities.
	
	Inspired by these biological processes, swarm robotics tries to emulate such complex behavior with a swarm of rather simple entities. Typically, these robots have basic movement and communication capabilities and can sense only parts of their environment, such as distances to other agents. Often, they are designed with small or even without memory systems, so that the agents can only access a short horizon of their perception. A common approach to imitate these systems is by extracting rules from the observed behavior. Kube et al \cite{Kube1998}, for example, investigate cooperative prey retrieval of ants to infer rules on how a swarm of robots can fulfill the task of cooperative box-pushing. Further work can be found in \cite{Martinoli2004}, \cite{Hoff2010}, \cite{Nouyan2009}. However, defining such behaviors manually can be tedious and the complexity of the tasks that we can solve manually is limited.
	
	In this paper, we follow a data-driven reinforcement learning approach to solving cooperative multi-agent tasks based on the locally sensed information of an agent. Such tasks are challenging learning problems since 1) the dimensionality of the problem grows exponentially with the number of agents and 2) the agents only partially observe a global state. However, in many scenarios, a fully observed global system state is available during training, for example, when a camera is filming the whole swarm. Using the global state information, we simplify the problem of evaluating the actions of the robots, similarly to a trainer who observes a team of football players and coordinates their actions and tactics during the training process. In the context of reinforcement learning, this mechanism can be formulated as an actor-critic learning algorithm \cite{Sutton1999}. While the actor learns a decentralized control policy operating on local observations, we provide the critic with the full system state to guide the swarm through the learning process.
	
	We propose a deep reinforcement learning framework to learn policies for a set of homogeneous agents who try to fulfill a cooperative task. We use a compact representation of the full system state, for example, the cartesian coordinates of all robots, to learn the Q-function while the actions of the agents are solely based on the local observations of each agent. Classical reinforcement learning algorithms heavily rely on the quality of the (typically hand-crafted) feature representation. Instead, our approach follows recent developments in deep reinforcement learning, which successfully combines techniques from deep learning with algorithms from reinforcement learning. Particularly appealing to this approach is the ability to learn policies in an end-to-end fashion, as mappings from high-dimensional sensory input to actions, without the need for complex feature engineering. Two examples are the deep Q-learning (DQN) algorithm \cite{Mnih2015} and the Deep Deterministic Policy Gradient (DDPG) algorithm \cite{Lillicrap2015}, which will be recapitulated in Section \ref{sec:bg}. There exist extensions to deal with partial observability \cite{Hausknecht2015}, \cite{Heess2015}, but, so far, they are only formulated for single-agent scenarios.
	
\begin{figure}
		\centering
		\includegraphics[height=4cm]{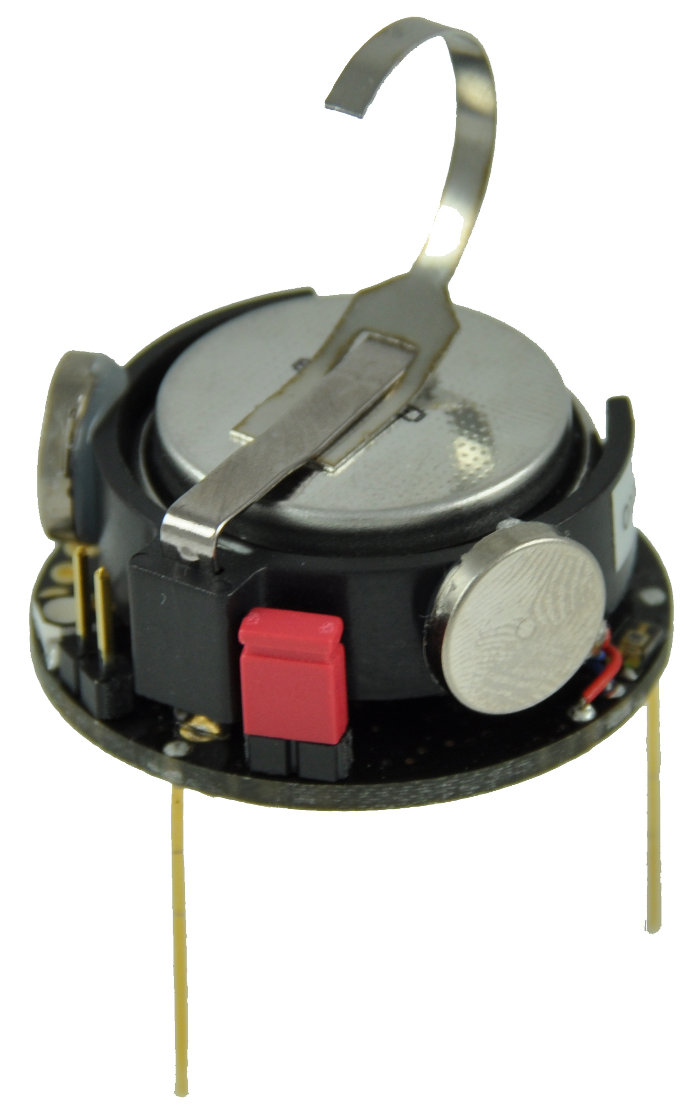}
		\caption{The figure shows a Kilobot robot. This platform is a base for the simulated agents in this paper.}
		\label{fig:kilobot}
	\end{figure}

	To demonstrate our framework, we formulate tasks in a simulated swarm environment, inspired by the Kilobot \cite{Rubenstein2014} robot platform (see Figure \ref{fig:kilobot}). The agents can move forward, turn left and right, and sense distances (but not the direction) to neighboring agents within a certain radius. Using only this information, we learn a policy which is able to 1)~establish and maintain a certain distance between agents and to 2)~cooperatively locate a target.
	
	\section{Background}
	\label{sec:bg}
	\noindent In this section, we summarize algorithms on which our method is based on. Two important single-agent reinforcement learning algorithms, the DQN algorithm and the DDPG algorithm, will be explained in the following.

	In single-agent reinforcement learning, a task can be formulated as a Markov decision process (MDP) which is a tuple $\langle \mathcal{S}, \mathcal{A}, P, R, \gamma \rangle$, where $\mathcal{S}$ is a set of states, ${\mathcal{A}}$ is a set of actions, $P: \mathcal{S} \times \mathcal{S} \times \mathcal{A} \rightarrow [0, \infty)$ is a transition density function, $R: \mathcal{S} \times \mathcal{A} \rightarrow \mathbb{R}$ is a reward function, and $\gamma \in [0, 1)$ is a discount factor. The discounted return $G_t^\gamma = \sum_{k=t}^{\infty} \gamma^{k-t} R(s_k, a_k)$ is the cumulative discounted reward the agent achieves. The Q-function is defined as the expected total discounted reward $Q(s, a) = \mathbb{E}[G_1^\gamma \mid s_1 = s,\, a_1 = a,\, \pi]$ following policy $\pi: \mathcal{S} \rightarrow \mathcal{A}$. The core problem is to find a policy for the agent which maximizes the expected return $J(\pi)=\mathbb{E}[G_1^\gamma \mid \pi]$. 
	
	\subsection{The Deep Q-Learning Algorithm}
	\noindent The DQN uses a deep neural network with parameters $\theta^Q$ to approximate a Q-function for continuous state and discrete action spaces. Q-learning \cite{Watkins1992} is applied to learn the policy. Techniques like a target network \cite{Mnih2015} and experience replay \cite{Lin1993} are used to make the update of the Q-function more stable for a deep neural network, which is a non-linear function approximator. The loss function
	\begin{align*}
	L(\theta^Q) &= \mathbb{E}_{e \sim \mathcal{D}} \left[\left(y - Q(s,a\mid \theta^Q)\right)^2\right],\\
	y &= r + \gamma \max_{a'} Q'(s',a'\mid \theta^{Q'})
	\end{align*}
	is used to update the parameters of the Q-function according to
	\begin{align*}
	\theta^Q &\leftarrow \theta^Q + \alpha_Q \nabla_{\theta^Q} L(\theta^Q), & \theta^{Q'} &\leftarrow \tau\theta^Q + (1 - \tau) \theta^{Q'},
	\end{align*}
	using, for example, an adaptive learning rate method like ADAM \cite{Kingma2014} and $\tau \ll 1$. Here, $\theta^{Q'}$ denotes the parameters of the target network which is a copy of the Q-function network, but updated at a much slower rate to ensure stability of the algorithm.
	The gradient of the loss function is given by
	\begin{align*}
	\nabla_{\theta^Q} L(\theta^Q) &= \mathbb{E}_{e \sim \mathcal{D}} \left[ \left(Q(s,a\mid \theta^Q) - y \right) \nabla_{\theta^Q} Q(s,a \mid \theta^Q) \right],\\
	y &= r + \gamma \max_{a'} Q'(s', a' \mid \theta^{Q'}),
	\end{align*}
	where experienced transitions $e=\langle s, a, r, s' \rangle$, stored in a buffer $\mathcal{D}$, are sampled in mini-batches for each update. Before learning begins, a warm-up phase is used to fill the replay memory with samples created from the initial exploration policy. The final policy is the greedy strategy $\pi(s) = \argmax_a Q(s, a \mid \theta^Q)$, while during learning, an epsilon greedy strategy is typically used for exploration.
	
	\subsection{The Deep Deterministic Policy Gradient Algorithm}
	\noindent The DDPG algorithm is an actor-critic learning algorithm for learning deterministic continuous control policies in single-agent environments, extending DQN to continuous action spaces. A Q-function $Q(s,a \mid \theta^Q)$ (the critic) and a policy $\pi(s \mid \theta^\pi)$ (the actor) are learned simultaneously, represented by neural networks with parameters $\theta^Q$ and $\theta^\pi$, respectively. Again, target networks $Q'$ and $\pi'$, whose parameters $\theta^{Q'}$ and $\theta^{\pi'}$ slowly track the parameters of the original networks, and experience replay is used. The update rules for the parameters are given by
	\begin{align*}
	\theta^Q &\leftarrow \theta^Q + \alpha_Q \nabla_{\theta^Q} L(\theta^Q), & \theta^{Q'} &\leftarrow \tau\theta^Q + (1 - \tau) \theta^{Q'},\\
	\theta^\pi &\leftarrow \theta^\pi + \alpha_\pi \nabla_{\theta^\pi}J(\pi), & \theta^{\pi'} &\leftarrow \tau\theta^\pi + (1 - \tau) \theta^{\pi'},
	\end{align*}
	where $J(\pi)=\mathbb{E}[Q(s,\pi(s|\theta^{\pi'}))]$.
	The gradients with respect to the critic parameters are given by
	\begin{align*}
	\nabla_{\theta^Q} L(\theta^Q) &= \mathbb{E}_{e \sim \mathcal{D}} \left[\left(Q(s,a \mid \theta^Q) -  y \right) \nabla_{\theta^Q} Q(s,a \mid \theta^Q) \right],\\
	y &= r + \gamma Q'\left(s', \pi'(s' \mid \theta^{\pi'}) \mid \theta^{Q'}\right),
	\end{align*}
	similar to the DQN algorithm. For the actor network, the update is given by
	\begin{align*}
	\nabla_{\theta^\pi} J(\pi) &= \mathbb{E}_{s\sim \mathcal{D}}\left[\nabla_{\theta^\pi} \pi(s\mid \theta^{\pi}) \nabla_a \left.Q(s, a)\right|_{a=\pi(s\mid \theta^{\pi})}\right],
	\end{align*}
	which is the result of the chain rule applied to the objective function $J(\pi)$ \cite{Silver2014}. Furthermore, we use the technique of inverting gradients introduced by \cite{Hausknecht2016} to bound the parameters of the action space.

	\section{Deep Reinforcement Learning\\ for Guided Swarm Learning}
	\noindent Our learning algorithm guides the learning process for homogeneous swarms by exploiting global state information which we assume to be available centrally during training. In swarm robotics, this information can be provided, for example, by a camera filming the scene. In our simulated experiments, we assume that the global state is given by all positions and orientations of the robots. 
	This information is not shared with the agents but used for evaluating the actions of the agents during the reinforcement learning process. Following a similar scheme as the DDPG algorithm, we learn a Q-function based on the global state information to evaluate the whole swarm's behavior. However, the actions chosen by the agents are determined based on observation histories only. This allows a guidance of the swarm during the learning process.
	
	\subsection{Problem Statement}
	\noindent We formulate the swarm system as a swarm MDP (see \cite{Sosic2016} for a similar definition) which can be seen as a special case of a decentralized partially observed Markov decision process (Dec-POMDP) \cite{Oliehoek2013}. An agent in the swarm MDP is defined as a tuple $\mathbb{A} = \langle \mathcal{S}, \mathcal{O}, \mathcal{A}, O\rangle$, where, $\mathcal{S}$ is a set of local states, $\mathcal{O}$ is the space of local observations, and ${\mathcal{A}}$ is a set of local actions for each agent. The observation model $O(o|s,i)$ defines the observation probabilities for agent $i$ given the global state. Note that the system is invariant to the order of the agents, i.e., given the same local state of two agents, the observation probabilities will be the same. The swarm MDP is then defined as $\langle M, \mathcal{E}, \mathbb{A}, P, R\rangle$, where $M$ is the number of agents, $\mathcal{E}$ is the global environment state consisting of all local states $\mathcal{S}^M $ of the agents and possibly of additional states of the environment,  and $P: \mathcal{S}^M \times \mathcal{S}^M \times \mathcal{A}^M \rightarrow [0, \infty)$ is the transition density function. 
	Each agent uses the same distributed policy $\mu: \mathcal{H} \rightarrow \mathcal{A}$, that maps histories $H$ of local observations and actions of the single agents to a new action to execute.
	We will denote $\eta$ as the horizon of the agent's history space, i.e., $\mathcal{H} = \mathcal{O}^\eta \times \mathcal{A}^\eta$. 
	In contrast to other Dec-POMDP solving algorithms and in accordance with the idea of swarm intelligence, we want to learn simple policies that do not explicitly tackle the problem of information gathering (i.e. nor reasoning about uncertainty of the state).
	All swarm agents are assumed to be identical. The reward function $R$ of the swarm MDP depends on the {\em global state} and all actions of the swarm agents, i.e., $R: \mathcal{S}^M \times \mathcal{A}^M \rightarrow \mathbb{R}$.
	
	\subsection{Centralized Guided Critic}
	\noindent For our guided learning approach, we require the action-value function $Q(s, a) = Q(s, \left[a^1, \dots, a^M\right])$ to evaluate the joint action $a \in \mathcal{A}^M$ of all agents in the current state. Tuples ${e = \langle s, h, a, r, s', h' \rangle}$ are stored in a replay memory and then redrawn for the updates, where $h$ denotes a joint history $[h^1,\dots, h^M]$ for a given time point. The gradients for the critic network parameters are given by
	\begin{align*}
	\nabla_{\theta^Q} L(\theta^Q) = \mathbb{E}_{e \sim \mathcal{D}} \left[\left(Q(s,a \mid \theta^Q) - y \right) \nabla_{\theta^Q} Q(s,a \mid \theta^Q) \right],
	\end{align*}
	where
	\begin{align*}
	y = r + \gamma Q'\left(s', \left[\mu(h'^1 \mid \theta^{\mu'}), \dots, \mu(h'^M \mid \theta^{\mu'}) \mid \theta^{Q'} \right] \right)
	\end{align*}
	are the target values defined by the target network and the target policy.
	The parameters of the target networks follow the set of original parameters with slow updates in a similar manner as in DDPG. Hence, the resulting algorithm for learning the Q-function is the same as used in DDPG with the difference that %
	a single policy function $\mu$ is used to compute the actions of all agents, whereas in DDPG, if we view the actions of all agents as a large combined action vector, the policy function would be different for each dimension of this action vector. Yet, the resulting actions of our guided critic approach can of course differ for the agents as the histories of the agents are different. 
	
	\subsection{Distributed Local Actors}
	\noindent In order to update the parameters of the actors, we formulate the deterministic policy gradient for swarm systems. The objective function $${J(\mu)=\mathbb{E}[G_1^\gamma \mid \mu]} \approx \mathbb{E}\left[Q\left(s, \mu(h^1 \mid \theta^\mu), \dots, \mu(h^M \mid \theta^\mu) \right) \right]$$ now depends on the actor in multiple ways as the actions of each agents are defined by the same actor function. Again, the gradient of the objective function is obtained by the chain rule, i.e., 
	\begin{align*}
	\nabla_{\theta^\mu} J =& \mathbb{E}_{e \sim \mathcal{D}} \left[ \sum_{i=1}^{M}\nabla_{\theta^\mu} \mu(h^i \mid \theta^\mu)\right. \dots \\
	& \left.\left.\nabla_{a^i} Q(s, [a^1, \dots, a^M] \mid \theta^Q)\right|_{a^i=\mu(h^i \mid \theta^\mu)} \right].
	\end{align*}
	We can see that the gradient is averaged over the gradients for the single actors. Given a specific global state $s$, the algorithm is able to improve the policy if the  histories in that state contain, in expectation, enough information to determine a high quality action. In the previous equation, this condition is formalized differentially, i.e., the policy improvement step is successful if the gradient of the actors is correlated with the gradient of the Q-function.   
	
	\subsection{Modeling Communication}
	\noindent In our swarm environment, the agents cannot access the global state. Instead, each agent is able to sense its local environment. Inspired by the Kilobot platform, in our scenarios the agent can sense distances to neighboring agents if they are within a certain range $d$. Naturally, the number of observed neighbored agents at each time step is changing over the course of an episode, resulting in a varying amount of observations, i.e., distance measurements, for each agent. Because we are dealing with neural networks, we are limited to a fixed input dimensionality of the observation representation. Thus, instead of collecting a vector of raw distance measurements at each time step, we use a histogram over distances with a fixed dimension. The histogram over distances for agent $i$ is given by the vector $u^i$, consisting of entries
	\begin{align*}
	u^i_{k} = \frac{\sum_{m=1}^{M} \bm{1}_{[0, d]} (d_m^i) \psi_k(d^i_m)}{\sum_{k'=1}^{K} \sum_{m=1}^{M} \bm{1}_{[0, d]} (d_m^i) \psi_{k'}(d^i_m)},
	\end{align*}
	where $d_m^i$ is the Euclidean distance between agent $i$ and agent $m$ and
	\begin{align*}
	\bm{1}_{[a,b]}(x) &= \begin{cases}
	1 & \text{if \,} x \in [a, b],\\
	0 & \text{else}
	\end{cases}
	\end{align*}
	is an indicator function. We use $K$ radial basis functions,
	\begin{align*}
	\psi_k(d^i_m )= \exp\left(-\frac{( d^i_m - \mu_k)^2}{2\sigma^2}\right), \ k=1,2, \dots, K,
	\end{align*}
	with $\sigma = 0.25$ and $\mu_0$ to $\mu_K$ equally spaced in the range of the communication radius.

	\section{Experiments}
	\label{sec:exp}
	
	\noindent In this paper, we use simulated swarms of agents whose capabilities are inspired by the Kilobot robots \cite{Rubenstein2014}.
	The Kilobot (Figure \ref{fig:kilobot}) is a three legged robot on a $\SI{33}{\milli\m}$ diameter circular base.
	It moves with the help of two vibrating motors which can be controlled independently.
	These motors let the robot move forward with a maximum speed of $\SI{1}{\cm\per\s}$ and turn with $\SI{45}{\degree\per\s}$.
	An ambient light sensor is placed on top of the robot. Two infrared sensors, one for transmitting and one for receiving messages, are placed beneath.
	Messages travel by reflection off the ground surface up to a distance of $\SI{7}{\centi\m}$, depending on the composition. It is also possible to determine the distance to the transmitting robot based on the intensity of the received message's signal.\\
	Videos of some of our learned policies can be found in a Dropbox folder \href{https://www.dropbox.com/sh/k7fzx5akx5jlwml/AAAfckZs7Tu3IU5Nb7WiruwRa?dl=0}{here}.
	
	\subsection{Simulation Environment}
	\noindent For the experimental results in this paper, we consider agents that move in a 2D world arranged as a torus. The global state $s = [s^1, \dots, s^M] \in \mathcal{S}^M$ is comprised of local states $s^i$ of each individual agent and an additional state of the environment, for example, the location of an object which the agents have to localize. Each agent's action is a two dimensional vector $a^i = [a_r^i, a_\phi^i] \in \mathcal{A}$, with $\mathcal{A} = \{[a_r, a_\phi] \in \mathbb{R}^2 : 0 \leq a_r \leq 1,\ |a_\phi| \leq \pi\}$, and the global action is given by $a = [a^1, \dots, a^M] \in \mathcal{A}^M$. The communication radius is chosen to be $d=4$ and the histogram over distances consists of $K=21$ entries.
	
	\begin{figure}
		\setlength\figureheight{0.5\textwidth}
		\setlength\figurewidth{0.5\textwidth}
		\centering
		\begin{subfigure}[b]{0.5\textwidth}
			\centering
			\includegraphics{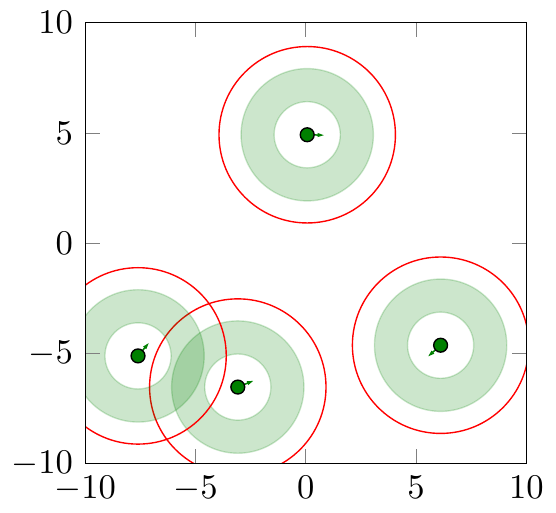}
			\caption{Graph Task}
			\label{fig:graph}
		\end{subfigure}
		\hspace{-2em} %
		\begin{subfigure}[b]{0.5\textwidth}
			\centering
			\includegraphics{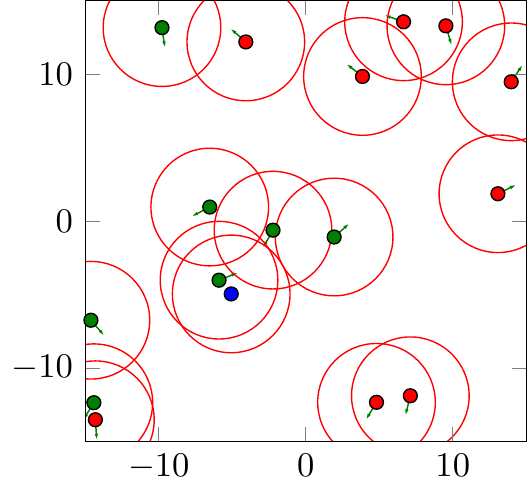}
			\caption{Localization Task}
			\label{fig:target}
		\end{subfigure}
		
		\caption{Visualization of the simulation environment. The left figure shows the graph task where agents are depicted with green dots and an orientation arrow. The outer red circle shows the communication radius and the inner light green area the distance in which a valid edge is established. The right figure shows the localization task. As long as agents have not found the target they are depicted by a red dot, after finding the target the color is changed to green. The outer red circle again shows the communication radius. The target is depicted by a blue dot.}
		\label{fig:sim_swarm}
	\end{figure}
	
	\subsection{Task 1: Building a Graph}
	\noindent In the first task, the agents have to find and maintain a certain distance to each other. Each agent is seen as a node in a graph. If there is a specified distance between two agents, we add an edge between these two nodes. The task is to maximize the number of edges.
	Figure \ref{fig:graph} shows a scene from our simulation with 4 agents. The agents are depicted as green dots with an arrow to show their orientation. The outer red circle indicates the communication radius and the inner light green ring the area in which another agent has the right distance such that a new edge is created. An agent's local state is given by the 2D position and orientation of the agent, i.e., $s^i = [x^i, y^i, \phi^i] \in \mathcal{S} = \{[x, y, \phi] \in \mathbb{R}^3 : |x| \leq 10,\ |y| \leq 10\}$. The next location and orientation of each agent is given deterministically by $s'^i = [(x^i + a_r^i  \cos(\phi^i+a_\phi^i) + 10) \mod 20 - 10, ~y^i + (a_r^i  \sin(\phi^i+a_\phi^i) + 10) \mod 20 - 10, ~\phi^i+a_\phi^i]$. 

	The agents are rewarded based on the number of pairwise distances between \num{1.5} and \num{3}. The reward function is given by
	\begin{align*}
	R(s, a) &= \sum_{i=1}^{M} \sum_{m>i}^{M} \bm{1}_{[1.5, 3]} (d_m^i) - 0.05 \sum_{i=1}^{M} \lVert a^i \rVert_2.
	\end{align*}
	It is designed to count only unique edges. 
	
	\subsection{Task 2: Localizing a Target}
	\noindent The second task requires the agents to cooperatively locate a target. Figure \ref{fig:target} shows a scene of the task with 16 agents. The agents are again depicted as green dots with an arrow to show their orientation and the outer red circle indicates the communication radius. Once an agent sees the target for the first time, it is depicted by a green dot. The target is depicted by a blue dot. An agent's local state is given by $s^i = [x^i, y^i, \phi^i, l^i] \in \mathcal{S} = \{[x, y, \phi, l] \in \mathbb{R}^3 : |x| \leq 15,\ |y| \leq 15, l \in \{0, 1\}\}$. The next location and orientation of each agent is given deterministically by $[x'^i, y'^i, \phi'^i] = [(x^i + a_r^i  \cos(\phi^i+a_\phi^i) + 15) \mod 30 - 15, ~y^i + (a_r^i  \sin(\phi^i+a_\phi^i) + 15) \mod 30 - 15, ~\phi^i+a_\phi^i]$. The localization bit $l^i$ is initialized with 0 and becomes 1 once the agent has found the target.
	The agents are provided with the observation
	\begin{align*}
	o^i = \begin{bmatrix} l^i, & d^i_T, & b^i, & u^i \end{bmatrix}
	\end{align*}
	where $l^i$ is the localization bit, $d^i_T$ is the distance to the target if it is within the communication radius (-1 if it is not), $b^i$ is a bit if a neighboring agent currently sees the target, and $u^i$ is the distribution over distances to agents within the communication range. The reward function is given by
	\begin{align*}
	R(s, a) &= \sum_{i=1}^{M} l^i - 0.05 \sum_{i=1}^{M} \lVert a^i \rVert_2
	\end{align*}
	and simply counts the number of agents who have already located the target with an additional action punishment.

	\subsection{Hyper Parameter Setup}
	\noindent We initialize the feed-forward neural networks with the following architectures and parameters. The critic has three and the actor four hidden layers with $[512, 256, 128]$ and $[1024, 512, 256, 128]$ neurons in the hidden layers, respectively. The input data to each layer is processed by a fully connected layer followed by exponential linear units as activation functions. The parameters of the hidden layers are initialized from a uniform distribution $[-\frac{1}{\sqrt{f}},\frac{1}{\sqrt{f}}]$ where $f$ is the dimensionality of the input of each layer. The output layer of the critic is also a fully connected layer with a linear activation function, initialized by a uniform distribution $[-3 \cdot 10^{-4}, 3 \cdot 10^{-4}]$, and the parameters of the output layer of the actor are initialized randomly from a uniform distribution $[-3 \cdot 10^{-3}, 3 \cdot 10^{-3}]$. The bounds of the outputs of the actor function $a^i_{r}$ and $a^i_{\phi}$ are enforced using the inverted gradient method \cite{Hausknecht2016}.
	
	We use a replay buffer of the last \num{500000} experience tuples and learn in mini-batches of size \num{32}. The base learning rate for the actor and the critic is chosen as \num{0.0001}. The decay rate for the soft target updates is $\tau = \num{0.0001}$.
	
	\begin{figure*}
		\setlength\figureheight{4cm}
		\setlength\figurewidth{4cm}
		\begin{subfigure}[b]{\textwidth}
			\centering
			\begin{subfigure}[b]{0.24\textwidth}
				\includegraphics{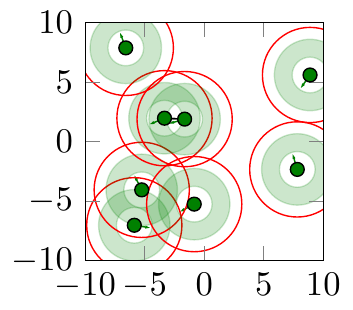}
				\caption*{$t=0$}
				\label{fig:8a_0}
			\end{subfigure}
			\begin{subfigure}[b]{0.24\textwidth}
				\includegraphics{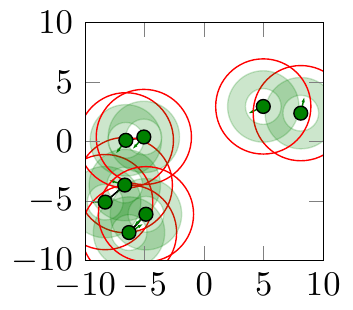}
				\caption*{$t=10$}
				\label{fig:8a_10}
			\end{subfigure}
			\begin{subfigure}[b]{0.24\textwidth}
				\includegraphics{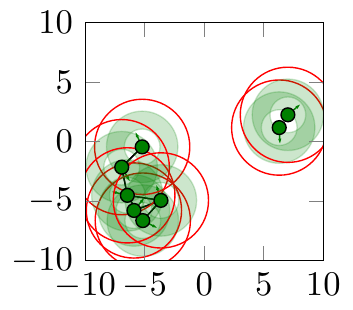}
				\caption*{$t=20$}
				\label{fig:8a_20}
			\end{subfigure}
			\begin{subfigure}[b]{0.24\textwidth}
				\includegraphics{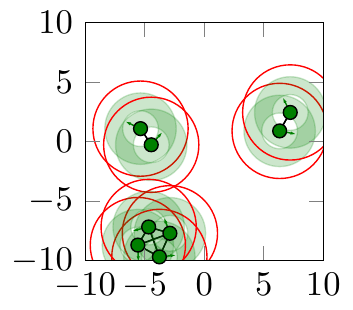}
				\caption*{$t=150$}
				\label{fig:8a_150}
			\end{subfigure}
			\caption{Policy learned by 3 agents.}
		\end{subfigure}
		
		\begin{subfigure}[b]{\textwidth}
			\centering
			\begin{subfigure}[b]{0.24\textwidth}
				\includegraphics{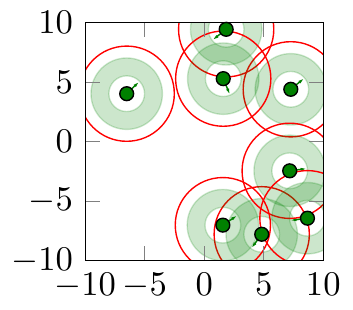}
				\caption*{$t=0$}
				\label{fig:6_on_8a_0}
			\end{subfigure}
			\begin{subfigure}[b]{0.24\textwidth}
				\includegraphics{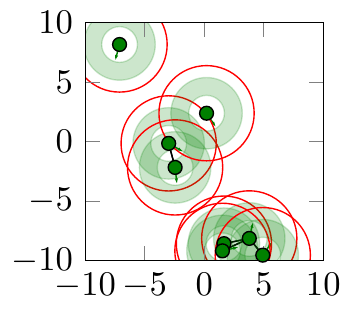}
				\caption*{$t=20$}
				\label{fig:6_on_8a_10}
			\end{subfigure}
			\begin{subfigure}[b]{0.24\textwidth}
				\includegraphics{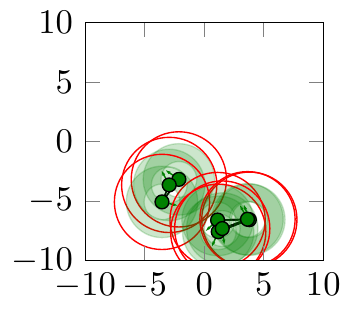}
				\caption*{$t=100$}
				\label{fig:6_on_8a_100}
			\end{subfigure}
			\begin{subfigure}[b]{0.24\textwidth}
				\includegraphics{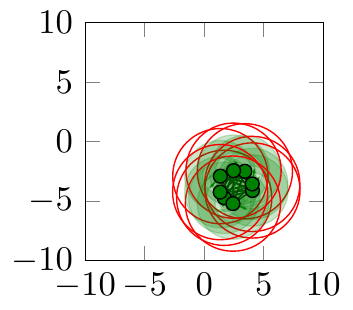}
				\caption*{$t=350$}
				\label{fig:6_on_8a_350}
			\end{subfigure}
			\caption{Policy learned by 6 agents.}
		\end{subfigure}

		\caption{Progression of an episode of 8 agents executing a policy learned by 3 and by 6 agents. In the beginning they are randomly placed in the scene. Over the course of the episode groups of agents are established. These groups move in circular patterns trying to keep their distance to each other.}
		\label{fig:8a}
	\end{figure*}
	
	\subsection{Results}
	\noindent For both tasks, we conducted experiments with \num{2}, \num{3}, \dots, and \num{8} agents learning simultaneously in the environment. For each scenario, we evaluated four independent training trials. More evaluations were unfortunately not possible due to the high computational demands of deep learning architectures. 
	One episode was of length $T=500$ time steps and the horizon of each agent's history was chosen to be $\eta = 10$ time steps. At the beginning of an episode, all agents were placed randomly in the world with random orientations. 
	
	\subsubsection{Task 1}
\begin{figure*}
		\setlength\figureheight{6cm}
		\setlength\figurewidth{12cm}
		\centering
		\includegraphics{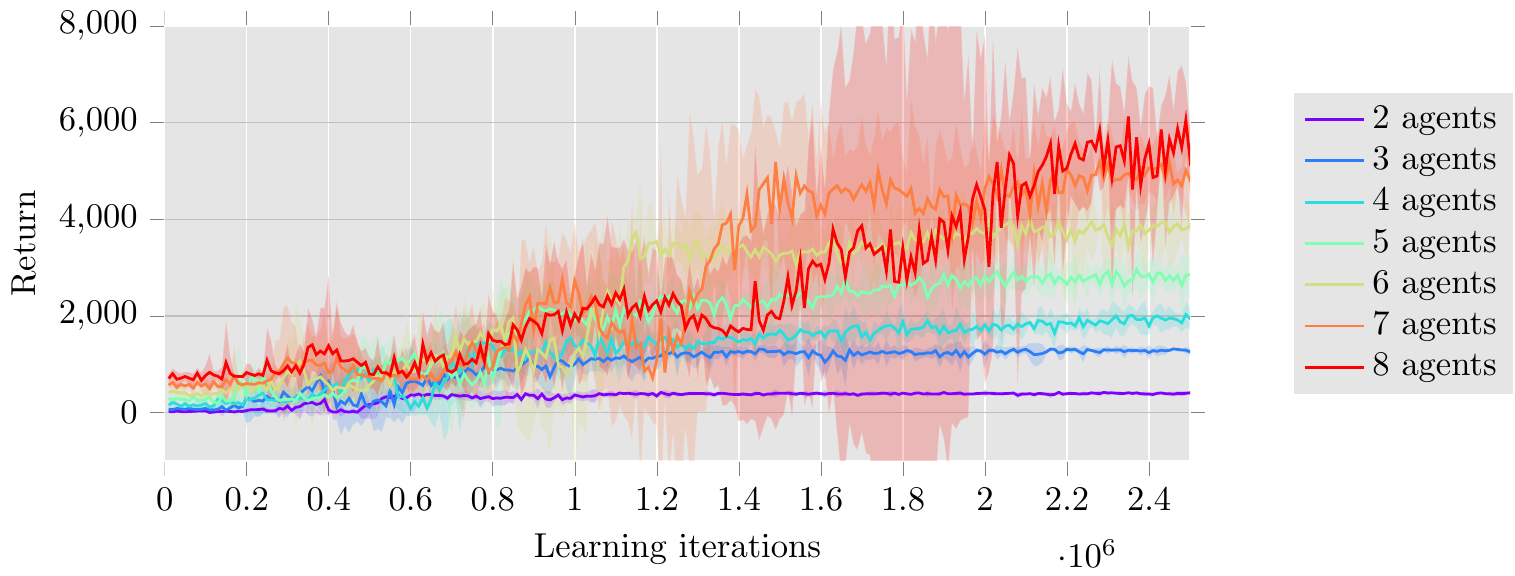}
		\caption{Mean learning curve for 2 to 8 agents with two times standard deviation for the edge task.}
		\label{fig:2_8_a_lc_model17c}
	\end{figure*}

	\noindent Figure \ref{fig:2_8_a_lc_model17c} shows mean learning curves where each policy was evaluated \num{50} times every \num{20} episodes during the learning process without exploration noise. Successful policies were learned in all scenarios, but, the more agents were part of the learning process, the harder the problem was to solve, resulting in noisier outcomes of the evaluation.
	
\begin{figure}
		\setlength\figureheight{5cm}
		\setlength\figurewidth{5.5cm}
		\centering
		\includegraphics{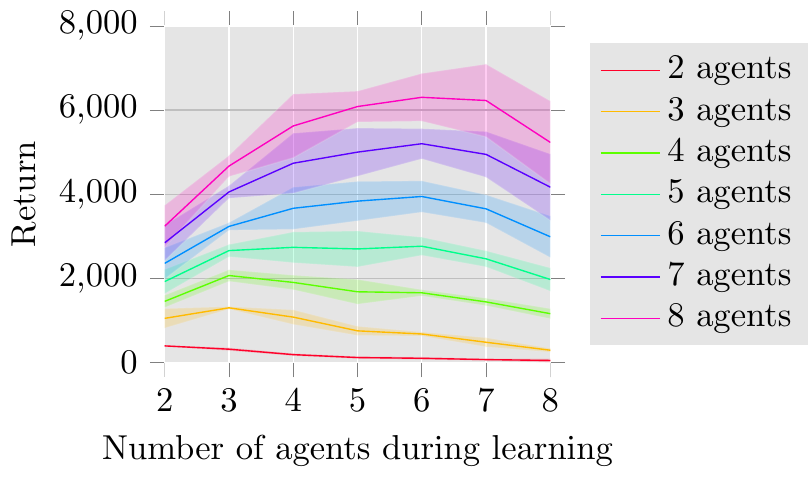}
		\caption{Evaluation of all learned policies of the edge task executed on 2 to 8 agents. Each policy is run 500 times and the plots show the mean return of the learned policies and two times the standard deviation.}
		\label{fig:eval_model17c}
	\end{figure}

	A cross evaluation using all learned policies with different configurations can be found in Figure \ref{fig:eval_model17c}. It shows the results of each policy of each scenario executed \num{500} times with the averaged return of each episode. We can see that policies learned by a low number of agents perform better on scenarios with few agents while policies learned by more agents perform better on scenarios with more agents. The policies learned on 7 and 8 agents do not reach the quality of the policy learned by 6 agents.
	
	Figure \ref{fig:8a} shows an example of a scenario with 8 agents at different time stages of an episode executing a policy learned by 3 and 6 agents, respectively. Valid edges between agents are indicated by a black line. The policy learned by 3 agents makes the agents collect in small groups, each moving in a circular pattern. The policy learned by 6 agents, however, tries to accumulate as many agents as possible into one group, resulting in more edges compared to smaller groups (6 edges at $t=150$ with the policy learned by 3 agents compared to around 20 edges at $t=350$ with the policy learned by 6 agents).

	\begin{figure*}
		\setlength\figureheight{6cm}
		\setlength\figurewidth{12cm}
		\centering
		\includegraphics{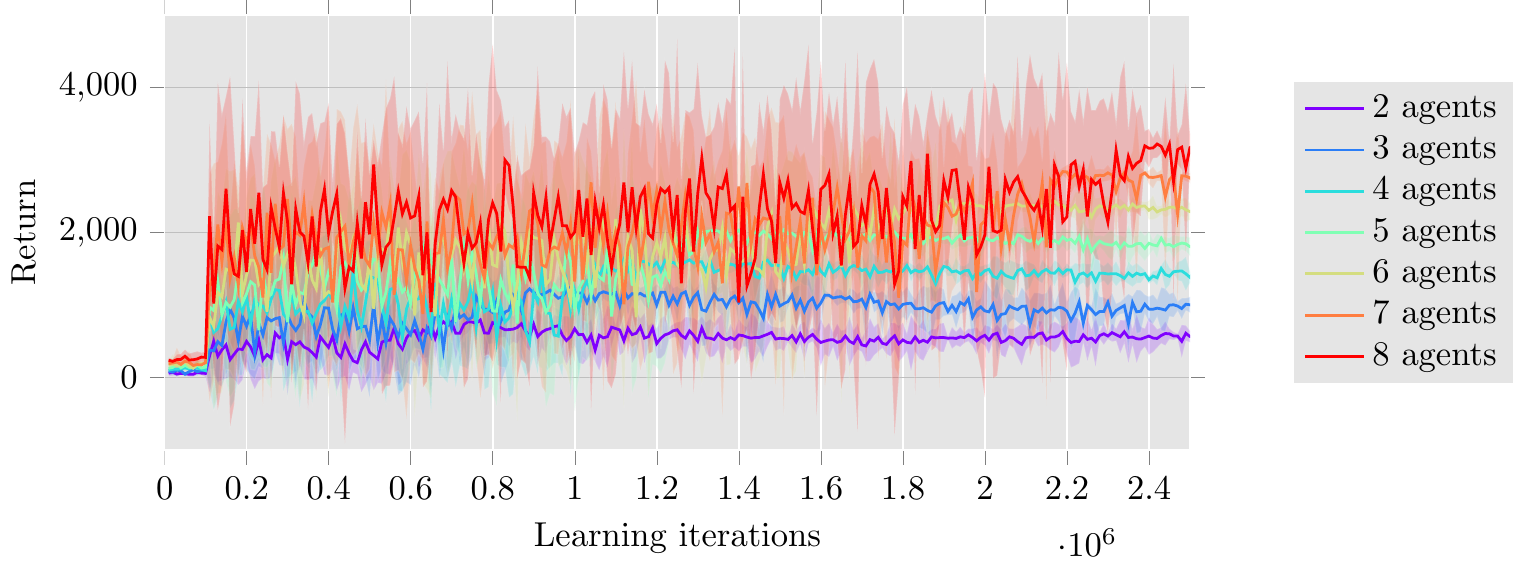}
		\caption{Mean learning curve for 2 to 8 agents with two times standard deviation for the localization task.}
		\label{fig:2_8_a_lc_model19}
	\end{figure*}
	
	\begin{figure*}
		\setlength\figureheight{6cm}
		\setlength\figurewidth{12cm}
		\centering
		\includegraphics{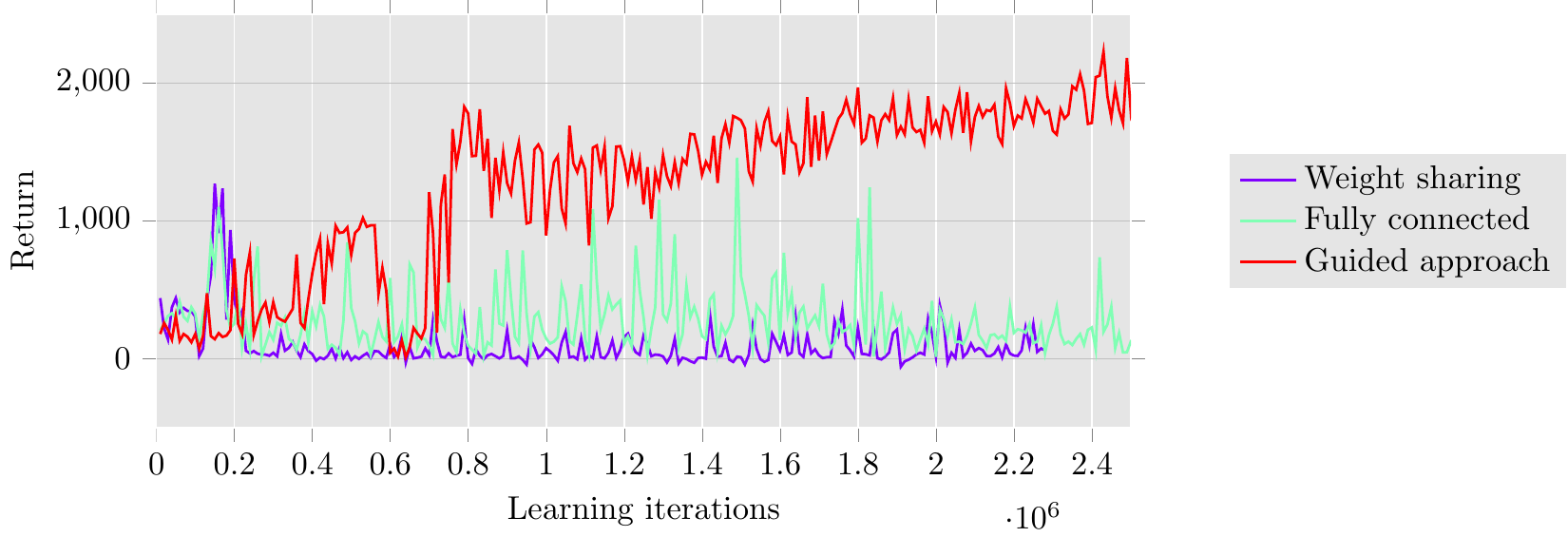}
		\caption{Learning curve of one experiment of the edge task with 4 agents using our guided approach and two exemplary learning curves using the non-guided approach.}
		\label{fig:lc_model2}
	\end{figure*}
	
	\subsubsection{Task 2}
\begin{figure}
		\setlength\figureheight{5cm}
		\setlength\figurewidth{5.5cm}
		
		\centering
		\includegraphics{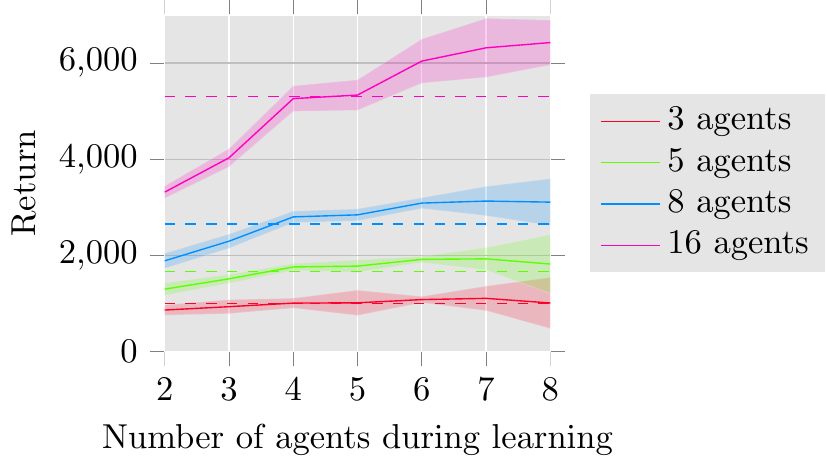}

		\caption{Evaluation of all learned policies of the localization task executed on 3, 5, 8 and 16 agents. Each policy is run 500 times and the plots show the mean return of all learned policies and two times the standard deviation. The dashed lines show the mean return of a policy without inter-agent communication.}
		\label{fig:eval_model19}
	\end{figure}

	\noindent For the second task, we conducted the same array of experiments. Figure \ref{fig:2_8_a_lc_model19} shows the learning curves of the localization task, again averaged over 4 trials each. Figure \ref{fig:eval_model19} shows an evaluation of all learned policies executed on different numbers of agents. Additionally, we learned a policy with one agent whose observation was if it has already seen the target, and the distance to the target if it is in its neighborhood (i.e. $\tilde{o}^i = \begin{bmatrix} l^i, & d^i_T \end{bmatrix}$). The mean return of this policy is drawn with dashed lines. From this comparison, we can see that policies learned by a low number of agents (typically less than 4) are not able to process the information they see on scenarios with more agents (which is reasonable given the fact that in a two agent scenario the agents hardly see each other). But, once enough agents take part in the learning process, the benefits of communication let these policies outperform the policy without inter-agent communication. The resulting behavior lets the agents who have already found the target stay in the vicinity of the target.
	
	\subsection{Guided Learning vs. Standard Learning}
	\noindent In order to show the efficacy of our approach, we tried to solve the edge building task with 4 agents using the joint action-observation history $h$ only, i.e. learning the Q-function $Q(h,a)$ instead of $Q(s,a)$.
	However, even with multiple different model architectures we were unable to learn meaningful policies. The architectures included variants with fully connected networks processing the joint histories, to weight sharing between the different agents' local histories, connecting later hidden layers to fully connected output layers, each using different numbers of hidden layers and neurons. Two exemplary learning curves are compared to a learning curve using the guided approach in Figure \ref{fig:lc_model2}.
	
	We suppose that two major factors play a role which lead to the failure of this approach. First, the dimensionality of the problem increases by a huge factor ($\dim(\mathcal{S}^4) = 12$ compared to $\dim(\mathcal{H}^4) = 920$). Second, the problem of learning a Q-function changes from a fully observed problem to a partially observed problem, where a history of length $\eta = 10$ contains too little information about the global state.
	
	\section{Conclusion}
	\label{sec:conc}
	\noindent In this paper, we presented the idea of guided policy learning for homogeneous swarm systems. We proposed an actor-critic learning approach and showed how a Q-function can be learned using global state information, while actors base their decisions on locally sensed information only. Using such an architecture, we could learn distributed control policies with limited sensing capabilities. Our results indicate that it is not feasible with a non-guided approach. We believe that guiding the learning process with global state information has the potential to scale current successes in deep reinforcement learning also to cooperative multi-agent scenarios such as robot swarms, which are beyond the reach with current methods. In many scenarios, more state information can be acquired during learning, for example, by using additional sensors.  

	\bibliographystyle{abbrv}
	\bibliography{llncs_arxiv}  %

\end{document}